# New Results and Long-standing Questions in Finite Temperature QCD

Frithjof Karsch[a]

[a] HLRZ, c/o KFA Jülich, D-52425 Jülich, Germany; and
Fakultät für Physik, Universität Bielefeld, P.O.Box 100131, D-33501 Bielefeld, Germany.

We review recent results on QCD at finite temperature. In particular we will discuss the chiral phase transition in two flavour QCD and new results on the excitation spectrum in the plasma phase of QCD.

## 1. INTRODUCTION

One of the most exciting predictions of QCD is the existence of a phase transition at some critical temperature to a new phase of strongly interacting matter – the quark gluon plasma (QGP). Lattice investigations of finite temperature QCD (FTQCD), which have been in progress for nearly 15 years, have contributed a lot to our understanding of the phase transition and the nature of the high temperature phase. The determination of the phase transition temperature, the order of the phase transition and temperature dependence of bulk thermodynamic quantities are central targets of lattice investigations at finite temperature. Nonetheless, the analysis of FTQCD still leaves us with many open questions. This is, for instance, the case for most aspects of the phase transition in the presence of light quarks, for the electric screening of the heavy quark potential in the QCD plasma phase and even more so for the magnetic sector of QCD at high temperatures. An understanding of the static as well as dynamic screening mechanisms in the high temperature phase is necessary in order to identify the fundamental excitation spectrum and quasi-particle structure of the QGP. All these questions have been addressed in lattice investigations over the past few years and the progress made, has been described in many excellent review talks at previous lattice conferences [1]. Still there are no final answers to the complex questions posed by FTQCD, and in this review we will again only be able to report on partial answers given by the numerous lattice studies performed during the past year.

In the case of QCD with dynamical fermions the quantitative results on bulk thermodynamic quantities are by far not as detailed as in the pure gauge sector where subtle aspects of the phase transition like the scaling behaviour of the latent heat or the surface tension can now be investigated [2,3]. The simulations with dynamical fermions on large lattices have mainly concentrated on an understanding of the nature of the phase transition and its flavour dependence. While it seems to be well established that the phase transition to the high temperature chiral symmetric phase of QCD is first order in the case of four light flavours [4] the situation is still unclear in the physically more interesting case of two light quark flavours. We will discuss the present status of these investigations in the next section.

A central goal of lattice investigations of the QGP is to understand the fundamental degrees of freedom in this phase and their interactions. Are these just weakly interacting quarks and gluons or do more complicated quasi-particle excitations or perhaps even bound states with hadronic quantum numbers exist? An answer to this question can come through an analysis of spectral functions in the QGP and will require the understanding of the analytic structure of various correlation functions. During the past few years this program has been pursued by examining, at finite temperature, the structure of correlation functions commonly used at zero temperature to study the properties of hadrons (masses, wave functions). The progress made in understanding the the analytic structure of such hadronic correlation functions as well as recent results for a related gluonic operator, the spatial string tension, will be discussed in section 3. In section 4 we give our conclusions.

## 2. TWO FLAVOUR QCD

The QCD phase transition in the presence of light quarks is expected to be controlled by the chiral symmetry properties of the QCD Lagrangian. In the limit of $n_f$ massless quarks the Lagrangian has a global $U_A(1) \times SU(n_f) \times SU(n_f)$ symmetry. At zero temperature the axial $U_A(1)$ is broken explicitly on the quantum level due to the non-vanishing anomaly while the chiral $SU(n_f) \times SU(n_f)$ flavour symmetry is broken spontaneously, giving rise to $(n_f^2 - 1)$ massless Goldstone particles. Universality arguments for a generic effective Lagrangian reflecting these symmetries suggest that for $n_f \geq 3$ the QCD phase transition is first order, while in the case of $n_f = 2$ the situation is less clear [5]. If the $U_A(1)$ symmetry is effectively restored at $T_c$, the transition is expected to be first order, while a second order transition can occur when the $U_A(1)$ symmetry only becomes realized at temperatures larger than $T_c$. The consequences of this latter scenario have been examined in some detail. Universal properties at the critical point are expected to be described by the critical exponents of the three dimensional $O(4)$ symmetry group [5,6] and an order parameter for the restoration of the $U_A(1)$ symmetry above $T_c$ has been constructed [7].

The analysis of the chiral phase transition in two flavour QCD on the lattice faces additional problems due the fact that neither in the staggered nor in the Wilson fermion formulation is the correct continuum chiral symmetry realized for finite lattice cut-off. The staggered fermion formulation does seem to have the advantage that at least a $U(1) \times U(1)$ subgroup of the continuum symmetry group is realized. This insures that even for a finite cut-off a symmetry restoring phase transition takes place. However, its properties might be influenced by the $U(1) \times U(1)$ symmetry as long as one does not get close enough to the continuum limit. This would lead to a second order phase transition with $O(2)$ critical indices, which has explicitly been shown to be the case in the strong coupling limit [8]. The fact, that at present even in the simulations on the large $8 \times 16^3$ and $8 \times 32^3$ lattices with rather small quark masses no first order phase transition has been found for the two flavour theory with staggered fermions [10,11] may thus not be the final answer for the two flavour theory in the continuum limit. A careful analysis of the critical behaviour is needed.

In the Wilson fermion formulation there is no continuous symmetry left and a priori one would thus not expect to find a finite temperature phase transition at finite lattice cut-off [1]. Nonetheless a complicated phase structure has been found in the Wilson formulation as well [9,12,13] and its relation to the finite temperature chiral transition has to be understood.

### 2.1. Staggered Fermions

The simulations of two flavour QCD on an $8 \times 16^3$ lattice with quark mass $ma = 0.004$, which have been performed by the Columbia group during the past year [11,18] gave some indications for metastabilities in the time histories of the order parameter and the plaquette expectation value at $\beta = 5.48$. This, however, did not find additional support in the new simulations on a larger lattice of size $8 \times 32^3$ at the same coupling, which have been presented at this conference [11]. These new data suggest that thermodynamic observables take on values which are between the values previously identified as upper and lower edge for possible discontinuity at a first order transition on the $8 \times 16^3$ lattice. Moreover, the average values calculated on the $8 \times 32^3$ lattice are in agreement with the average values extracted from the runs on the smaller lattice, if one averages over the whole data sample. This seems to rule out a first order phase transition. It should, however, be noted that the currently existing runs of the Columbia group on the large $8 \times 32^3$ lattice are still rather short and the above conclusions thus have to be considered as being preliminary.

[1]We here talk about a true phase transition with non-analytic behaviour of thermodynamic quantities in the limit of infinite spatial volume. It should be stressed that such a transition exists for staggered fermions in the limit of vanishing quark mass and should not be confused with the pseudo-critical behaviour found in thermodynamic quantities for non vanishing values of the quark mass.

| | $6/g_c^2(ma, N_\tau)$ | | |
|---|---|---|---|
| $ma$ | $N_\tau = 4$ | $N_\tau = 6$ | $N_\tau = 8$ |
| 0.004 | – | – | 5.450 (30) [18,21] |
| 0.00625 | – | – | 5.475 (25) [10,21] |
| 0.01 | 5.265 (5) [20] | – | – |
| 0.0125 | 5.270 (2) [19] | 5.420 (10) [15] | 5.490 (30)[16,17] |
| 0.025 | 5.288 (2) [19] | 5.445 (5) [15] | – |
| 0.025 | 5.291 (1) [20] | – | – |
| 0.05 | 5.320 (10) [14] | 5.470 (40)[14] | – |
| 0.1 | 5.375 (10) [14] | 5.525 (40) [14] | – |
| 0.0 | 5.225(5) | 5.363(4) | 5.424(16) |

Table: Pseudo-critical couplings on various size lattices and for various values of the bare quark mass [10,14–20]. The numbers for $N_\tau = 8$ are based on an analysis of existing data for the chiral order parameter $\langle\bar{\psi}\psi\rangle$ as discussed in Ref. [21]. The last row gives the results from the fits shown in Fig. 2

At present there are no indications for a first order transition in simulations with staggered fermions, which so far have been performed on lattice of size $N_\tau \times N_\sigma^3$ with temporal extent $N_\tau =$ 4, 6, 8 and various values of the quark masses $ma$. On these lattices pseudo critical couplings $\beta_c(N_\tau, ma)$ have been determined, which characterize the point of largest variation of the chiral order parameter, $\langle\bar{\chi}\chi\rangle$, and/or the Polyakov loop expectation value. In the table we have collected the results for all pseudo critical couplings determined so far for different parameter sets. As far as their dependence on the lattice size could be checked no systematic dependence has been found [19,21]. The results for the chiral order parameter on lattices of size $8 \times N_\sigma^3$ and three values of the quark mass are shown in Fig. 1. For $ma = 0.0125$ this illustrates that there is no visible difference between the results obtained on lattices with spatial extent $N_\sigma = 12$ and 16.

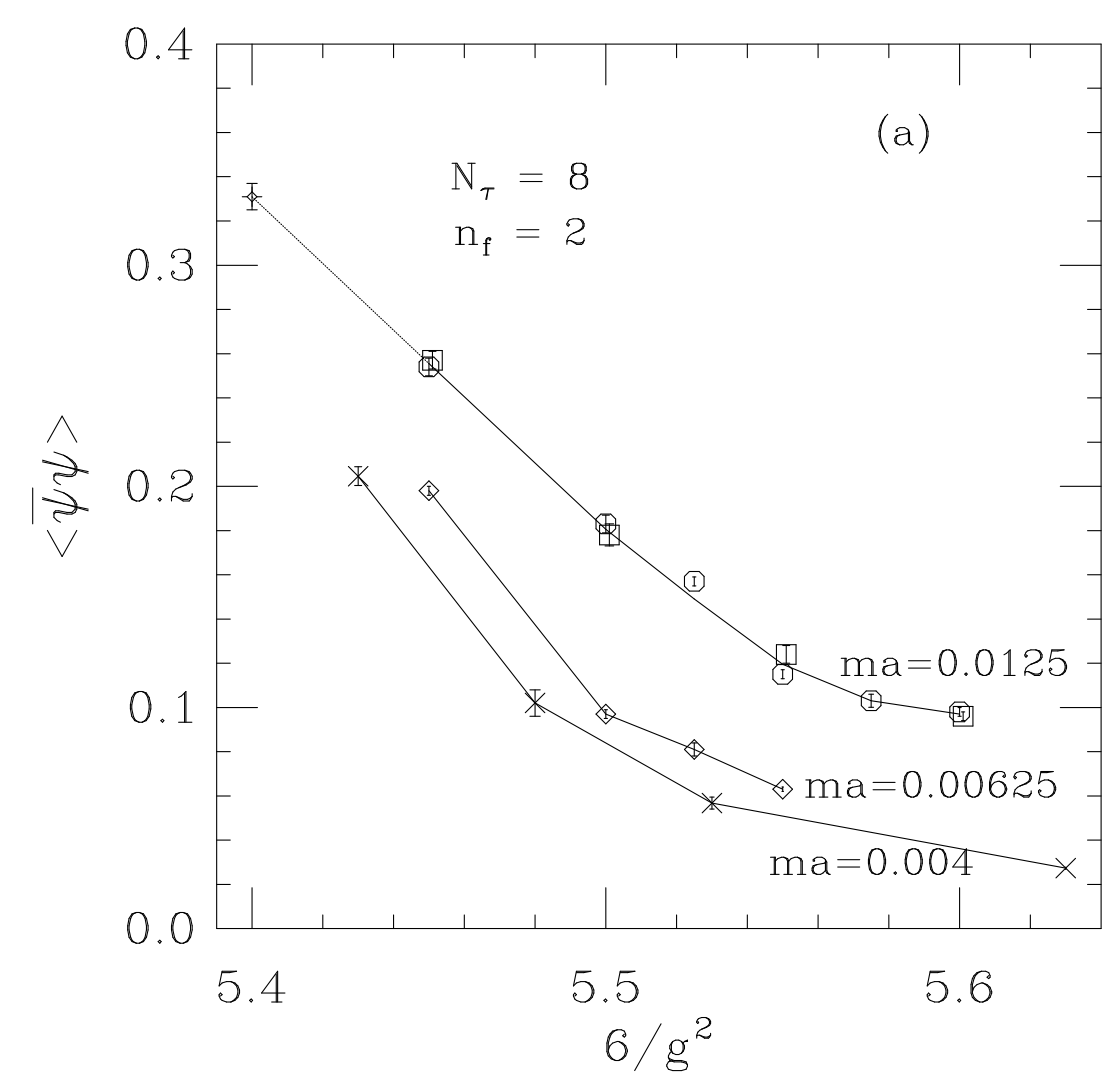

Figure 1. The chiral order parameter on lattices of size $8 \times N_\sigma^3$ and quark masses $ma = 0.0125$ [16,17], $ma = 0.00625$ [10] and $ma = 0.004$ [18]. The results shown are for $N_\sigma = 16$, except for the squares, which correspond to $N_\sigma = 12$.

Can we understand the quark mass dependence of the pseudo critical couplings in terms of scaling properties in the vicinity of a second order phase transition? Is the quark mass dependence compatible with the behaviour expected for a system with critical indices in the universality class of the three dimensional $O(N)$ spin models? In the presence of a symmetry breaking term in the action, as is the case for QCD with a non vanishing quark mass, finite size scaling techniques are of little help to answer these questions. For sufficiently large lattices the correlation length in the system will be limited by the non vanishing quark masses rather than the size of the system. In order to study the scaling properties at the chiral phase transition of QCD with dynamical light quarks one thus has to follow an approach, which differs from the finite size scaling tests used in the case of the deconfinement transition in pure $SU(N)$ gauge theories. We have to study the scaling properties in terms of the symmetry breaking fields (quark masses) and determine the scaling behaviour in the limit of vanishing quark mass [6,8,21]. The scaling properties can be deduced from the scaling behaviour of the singular part of the free energy

$$f_s(t,h) = b^{-1} f_s(b^{y_t} t, b^{y_h} h) \quad , \tag{1}$$

where $b$ is an arbitrary scale parameter, $t = (T - T_c)/T$ denotes the reduced temperature and $h$ is the coupling to the symmetry breaking field. The critical exponents $y_t$ and $y_h$ are related to the more familiar exponents $\beta$ and $\delta$ through

$$\begin{aligned} \beta &= \frac{y_h}{1-y_h} = \begin{cases} 4.755(6)\ , & O(2) \\ 4.82(5)\ , & O(4) \end{cases} \quad , \\ \delta &= \frac{1-y_h}{y_t} = \begin{cases} 0.3510(2)\ , & O(2) \\ 0.38(1)\ , & O(4) \end{cases} \quad . \end{aligned} \tag{2}$$

In the continuum limit of lattice QCD the couplings $t$ and $h$ can be expressed in terms of the couplings appearing in the lattice Lagrangian,

$$t = \frac{6}{g^2} - \frac{6}{g_c^2(N_\tau,0)} \quad , \quad h = maN_\tau \equiv m_q/T \ . \tag{3}$$

The second derivatives of the free energy give the singular part of susceptibilities. Making use of the scaling properties of $f_s$ one finds for the scaling behaviour of the location of the peak of the susceptibilities [21]

$$\frac{6}{g_c^2(N_\tau,ma)} = \frac{6}{g_c^2(N_\tau,0)} + c(maN_\tau)^{1/\beta\delta} \quad . \tag{4}$$

Eq. 4 can be used to test the scaling behaviour of the pseudo critical couplings given in the table, although it should be stressed that in most cases these couplings have not been extracted directly through a calculation of susceptibilities but rather through an examination of the slope of the chiral order parameter. The result of such a comparison is shown in Fig.2. A fit to the data set for the lattices with temporal extent $N_\tau = 4$ yields for the relevant combination of critical exponents [21]

$$\frac{1}{\beta\delta} = 0.57 \pm 0.16 \quad , \tag{5}$$

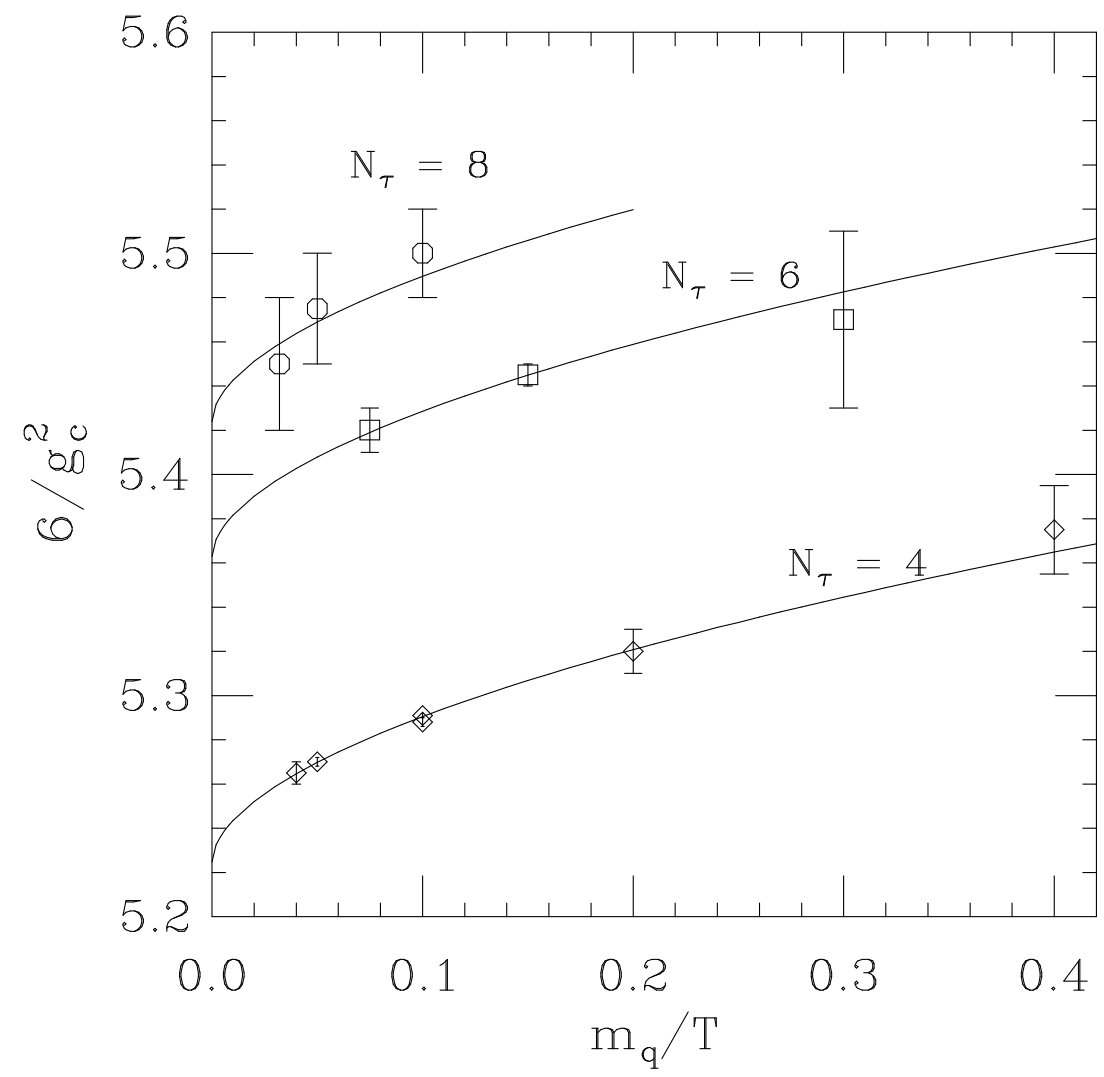

Figure 2. The pseudo critical couplings for two flavour QCD as function of $h = maN_\tau$ and temporal lattice sizes $N_\tau = 4$, 6 and 8.

which should be compared with the corresponding values 0.599 and 0.546 for the $O(2)$ and $O(4)$ symmetric models, respectively. Clearly, the presently available data do not allow one to distinguish between the two cases. However, even more remarkable than the proximity to the $O(N)$ exponents is the fact that the shift of the pseudo critical couplings does indeed seem to depend on the dimensionless coupling $h = maN_\tau$ alone rather than on $N_\tau$ and $ma$ separately as it would be the case in the strong coupling limit.

The presently available data for two flavour QCD in the staggered formulation are consistent with the behaviour expected in the case of a second order phase transition with $O(2)$ or $O(4)$ critical indices. Moreover, there is no indication for a substantial difference in the critical behaviour between lattices with temporal extent $N_\tau = 8$, and the smaller $N_\tau = 4$ lattices.

### 2.2. Wilson Fermions

The analysis of the chiral phase transition with Wilson fermions is not as straightforward as it is with staggered fermions. To start with, it is not obvious that QCD with Wilson fermions will have a finite temperature phase transition on lat-tices with finite temporal extent $N_\tau$. There is no global symmetry, which could be restored at high temperature, and thus one would not expect to find a second order phase transition in the infinite spatial volume limit of the lattice model. This situation is quite different from pure gauge theories or QCD with staggered fermions, where the existing global symmetries can lead to second order phase transitions at non vanishing lattice cut-off. If there is a finite temperature phase transition in the Wilson formulation at finite cut-off at all, one would thus expect to find only first order phase transitions.

The strategy in studying the QCD finite temperature phase transition with Wilson fermions usually consists of two steps, which are analogous to the procedure used in the staggered fermion formulation:

- On a lattice with given temporal extent $N_\tau$ one identifies the pseudo critical values of the hopping parameter, $\kappa_{pc}(N_\tau, g^2)$, at which a crossover from low temperature thermodynamics to that of the high temperature phase can be observed. This may be done by studying the behaviour of the Polyakov loop, the pion screening mass extracted from spatial correlation functions, or even by inspecting the behaviour of the plaquette expectation value.

- One tries to follow this pseudo critical line to the point where $\kappa_{pc}$ reaches $\kappa_c$. The latter being defined as the zero temperature critical line, $\kappa_c(g^2)$, where the pion mass vanishes at non vanishing lattice spacing.

While the first step is pretty obvious, it is less clear whether the second step can be realized. What is the relevance of $\kappa_c$ for the thermodynamics on lattices with finite temporal extent, $N_\tau$? Does it correspond to a critical point, which correponds to a phase transition point with vanishing quark masses? Does a coupling $6/g_c^2$ exist, where for a given value of $N_\tau$ we find $\kappa_{pc} = \kappa_c$?

The reason for the above questions already becomes evident in the non-interacting ($6/g^2 \to \infty$) weak coupling limit of lattice QCD with Wilson fermions. In this case the partition function on a $N_\tau \times N_\sigma^3$ lattice is given by

$$
\begin{aligned}
Z(\kappa) \quad = \quad & \prod_{q_0,\vec{q}} \Big[ \big(1 - 8\kappa + 4\kappa \sin^2(q_0/2) \\
& +4\kappa \sum_{\mu=1}^{3} \sin^2(q_\mu/2)\big)^2 \\
& +4\kappa^2 \sin^2 q_0 + 4\kappa^2 \sum_{\mu=1}^{3} \sin^2 q_\mu \Big]^2 \quad , \qquad (6)
\end{aligned}
$$

where the spatial momenta $q_\mu$ are even multiples of $\pi/N_\sigma$ and $q_0$ is non-zero and an odd multiple of $\pi/N_\tau$ due to the anti-periodic boundary conditions. For this reason one finds that even in the limit $N_\sigma \to \infty$ the zeroes of the partition function remain in the complex plane. The zeroes closest to the origin are given by

$$
\kappa_{pc}(N_\tau) = \frac{8 - \sin^2(\pi/2N_\tau) \pm i \sin(\pi/N_\tau)}{\big(8 - \sin^2(\pi/2N_\tau)\big)^2 + \sin^2(\pi/N_\tau)} \qquad (7)
$$

and lead, in the limit $N_\tau \to \infty$, to the familiar result $\kappa_c = 1/8$. The free Wilson fermion theory thus has a singular point (zero of the partition function on the real axis) with a diverging correlation length only in the limit $N_\tau \to \infty$. For finite $N_\tau$, however, Eq. 7 defines only a set of pseudo critical couplings. In view of this it seems to be unclear whether a $\kappa_c(g^2)$, at which, for instance, the pion screening mass (extracted from spatial correlation functions) becomes zero, exists for finite $N_\tau$ and non-zero $g^2$.

The first simulations in which the finite temperature transition with Wilson fermions has been studied [22] showed that the line of pseudo critical values $\kappa_{pc}(N_\tau, g^2)$ shifts into the strong coupling regime even for moderate values of the pion mass. The possibility has been raised that it would extend all the way down to $6/g^2 = 0$ without leading to small values of the pion mass at the crossover point. The line of pseudo critical $\kappa$-values has been examined by now in great detail for the case $N_\tau = 4$ [12,13,23,24]. In Fig. 3 we show the result of a systematic study of the crossover behaviour in the Polyakov loop on lattices of size $4 \times 8^3$ which has been performed by the MILC collaboration [13]. It can clearly be seen that the crossover is rather flat for large and small values of $6/g^2$ while there is a rather rapid crossover around $6/g^2 = 4.9$. A similar behaviour has been found for other observables.

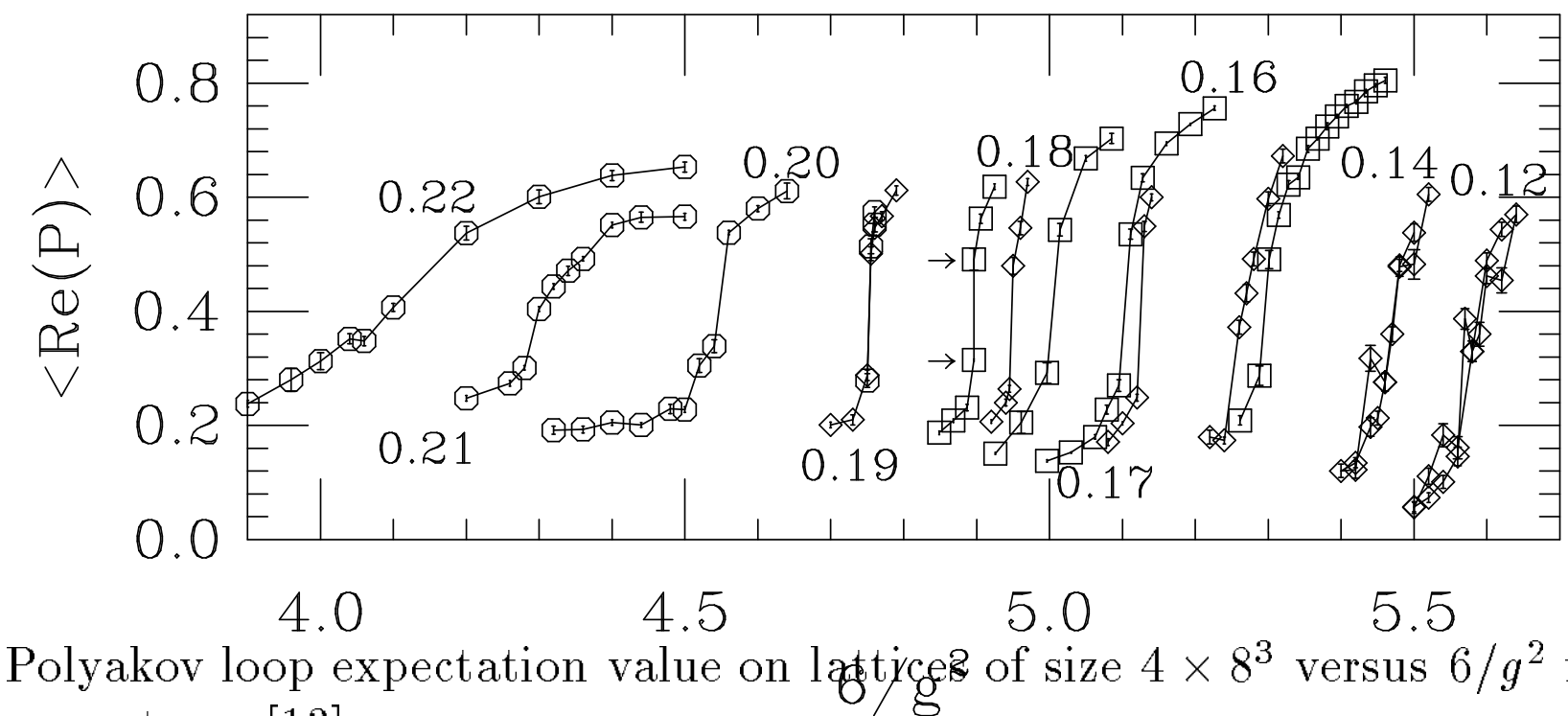

Figure 3. Polyakov loop expectation value on lattices of size $4 \times 8^3$ versus $6/g^2$ for various values of the hopping parameter $\kappa$ [13].

The QCDPAX collaboration has analyzed the behaviour of the pion screening mass on the crossover line. They find that this quantity is consistent with zero below $6/g^2 = (3.9 - 4.0)$ and that the value determined for the pseudo critical hopping parameter $\kappa_{pc}(4, g^2)$ in this region is consistent with the value $\kappa_c$ determined on larger lattices from the vanishing of the pion mass [23,24].

In Fig. 4 we show a collection of values of the pseudo critical hopping parameter $\kappa_{pc}$ on lattices with temporal extent $N_\tau = 4$ and 6 and compare them with data for $\kappa_c$. A collection of pion screening masses calculated from spatial correlation functions at $\kappa_{pc}$ as well as zero temperature pion masses calculated at the same $\kappa$ value is shown in Fig. 5. We note that this does not seem to suggest a vanishing of the screening masses in the vicinity of $6/g^2 = 4.0$, although $\kappa_{pc}$ clearly comes close to $\kappa_c$ in this regime. In any case we do not seem to have any indication of a *true* phase transition in this regime.

While the general structure of the phase diagram found for Wilson fermions on lattices with $N_\tau = 4$ seems to confirm our expectation that there will be no genuine finite temperature phase transition on lattices with finite temporal extent, the situation is complicated by the results found in simulations for $N_\tau = 6$ [25,26]. In this case a strong first order phase transition has been observed for gauge couplings smaller than $6/g^2 \simeq 5.0$. In fact, the results for $N_\tau = 6$ are drastically different from those of the $N_\tau = 4$ simulations in this region. Some results for plaquette expectation values are shown in Fig. 6. The differences in plaquette expectation values for $N_\tau = 4$ and 6 as well as the gap at the first order transition for $N_\tau = 6$ (at $\beta = 5.02$ for $\kappa = 0.18$) is much larger than what we know from simulations in the pure gauge sector or with staggered fermions. This might be due to the fact that in the Wilson fermion case the simulations are performed rather deep in the strong coupling regime. However, there also is another possibility, which at present cannot be ruled out: The large gap in the plaquette expectation values reminds one very strongly about the results obtained even deeper in the strong coupling regime in simulations with large number of flavours [9,24]. In the case of $n_f \geq 6$ strong first order phase transitions have been observed for finite $N_\tau$, which actually turned out to be temperature ($N_\tau$) independent. At present we cannot rule out that the first order transitions observed in the $n_f = 2$ case on $N_\tau = 6$ lattices are also temperature independent. This will certainly soon be clarified through simulations with larger values of $N_\tau$.

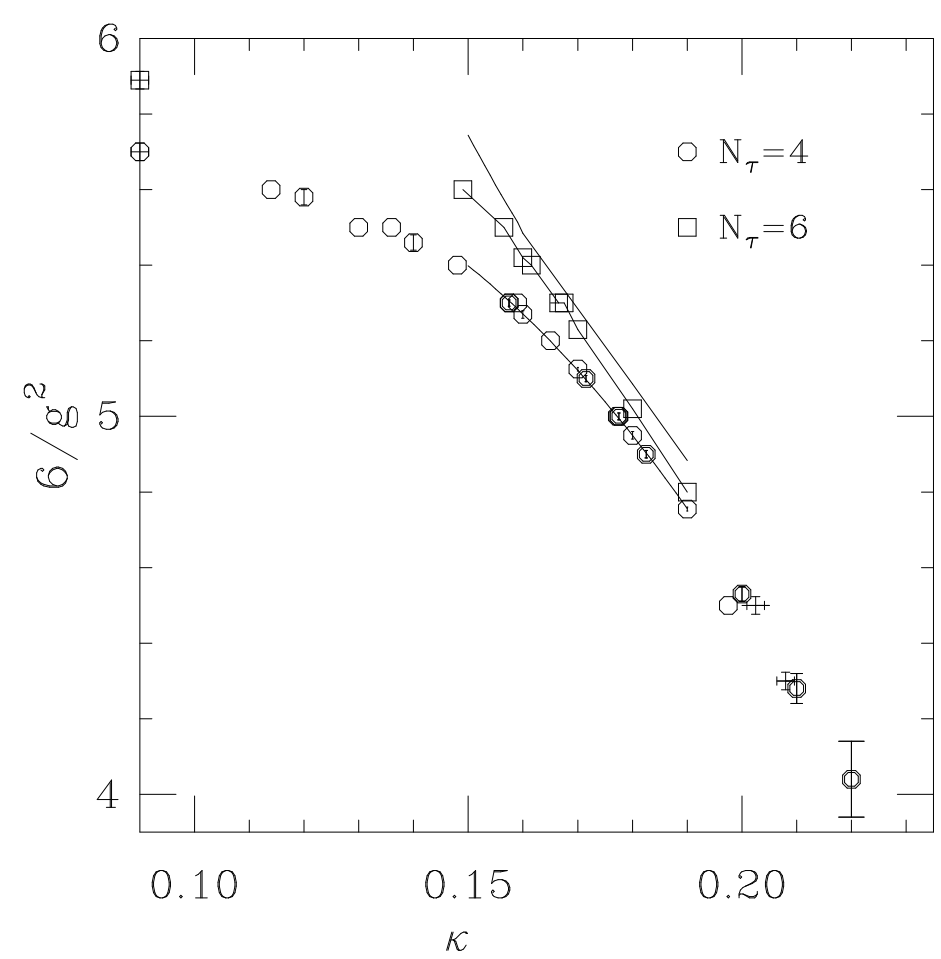

Figure 4. Pseudo critical hopping parameters determined on $N_\tau = 4$ and 6 lattices. The upper curve and the two fancy crosses indicate values for $\kappa_c$ determined at zero temperature.

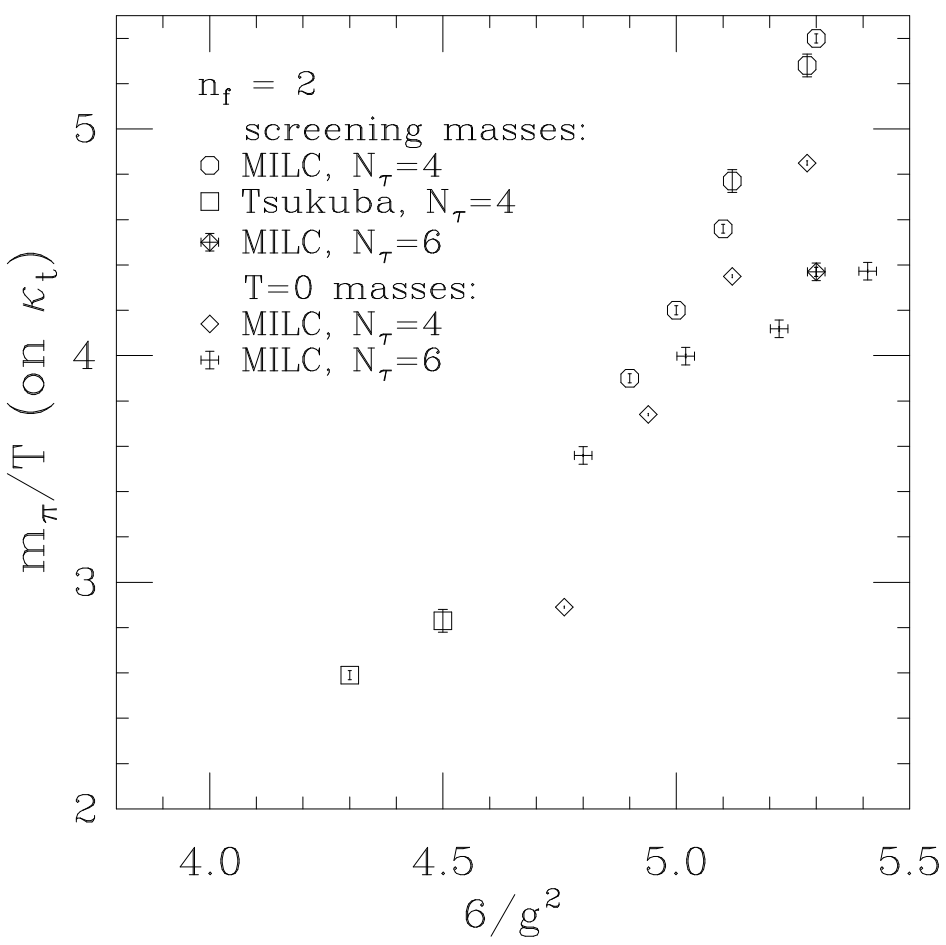

Figure 5. Screening masses extracted from spatial correlation functions on lattices with temporal extent $N_\tau = 4$ and 6 and zero temperature pion masses determined on lattices with large temporal extent. In both cases calculations have been performed at (or extrapolated to) the pseudo critical hopping parameter value $\kappa_{pc}(N_\tau, 6/g^2)$.

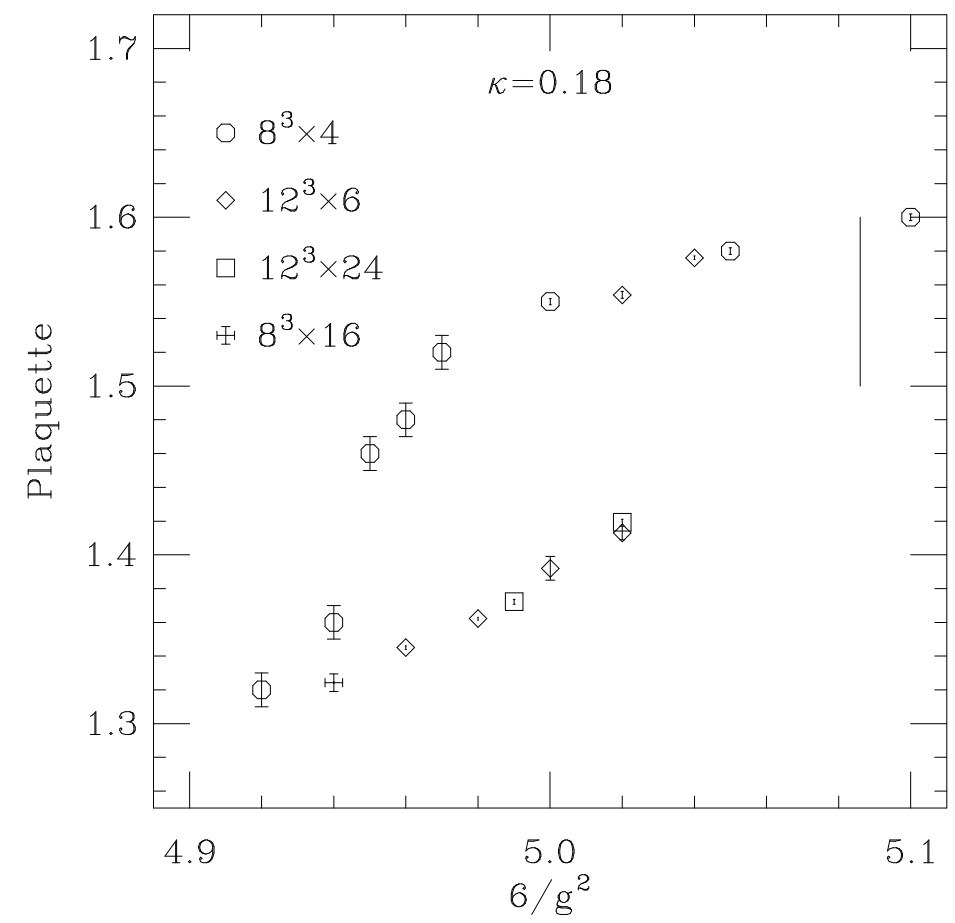

Figure 6. Plaquette expectation values versus $6/g^2$ on various size lattices for $\kappa = 0.18$. In the low temperature phase the plaquette expectation values on $N_\tau = 6$ lattices agree with the zero temperature results up to the first order transition at $\beta = 5.02$. The vertical line indicates the location of $\kappa_c$.

In the present simulations on lattices with $N_\tau = 4$ and 6 the crossover behaviour as well as the first order transitions occur at $\kappa$-values at which the zero temperature pion masses are still quite large, i.e. $m_\pi/T \gtrsim 4.0$. Following the line of pseudo critical couplings into the strong coupling regime does not seem to improve the situation. In fact, when comparing the zero temperature pion masses calculated at $\kappa_{pc}(N_\tau, g^2)$ for $N_\tau = 4$ and 6 it seems that below $6/g^2 \simeq 5.0$ the pion mass in units of the temperature increases with increasing $N_\tau$. This is also shown in Fig. 5. It thus seems that one has to concentrate on the coupling regime $6/g^2 > 5.0$ for the analysis of the finite temperature phase transition with Wilson fermions, if one tries to reach a regime where the pion mass is small in units of the temperature. Many of the peculiar results obtained with Wilson fermions so far seem to be due to specific features of the Wilson action in the strong coupling regime.

It is the hope that the situation will improve closer to the continuum limit. This, however, will require simulations with quite large values of $N_\tau$.

## 3. THE HIGH TEMPERATURE PHASE

Despite many years of analytical and numerical studies of finite temperature QCD little quantitative is known about the relevant excitations characterizing the structure of hot QCD in the vicinity of the phase transition to the QGP as well as in the high temperature phase itself. There are various indications that even at high temperatures the excitation spectrum might be complicated in particular due to non-perturbative effects in the magnetic sector of QCD. In order to make progress in our understanding of such non-perturbative effects the analytic structure of correlation functions has been studied in much detail over the last years.

At finite temperature the temporal extent of the lattice is limited to the interval $0 \leq \tau \leq 1/T$ and thus correlation functions cannot be studied at large Euclidean time separations. Hence the studies focused on the analysis of long distance properties of spatial correlation function. A comparison of the numerical results with model calculations and high temperature perturbation theory then allows one to extract information on the excitation spectrum.

During the past few years the structure of spatial correlation functions with hadronic quantum numbers has been examined in detail [27]. These studies have now also been extended to the analysis of hadron propagators in temporal direction [28,29] and glueball operators [31]. A related correlator, wich actually has been studied several years ago [32], is the spatial Wilson loop. This simulation confirmed the prediction of an area law behaviour for spatial Wilson loops above $T_c$ [33]. However, the detailed properties of the spatial string tension have only been examined this year [34–36]. We will discuss these new results in the next two subsections.

### 3.1. The Spatial String Tension

At high temperature an effective, three dimensional action can be derived by systematically integrating out contributions from non-static modes in the QCD Lagrangian. The effective theory is a three dimensional Gauge-Higgs model with an adjoint Higgs fields resulting from the static temporal component of the gauge potential. The effective action takes on the form

$$S_{\rm eff} = \frac{2N_c}{g_3^2}\left(S_w + S_h + S_v\right) \ , \ g_3^2 = g^2(T)T \qquad (8)$$

where $S_w$ denotes the action of a three dimensional $SU(N_c)$ gauge theory, $S_h$ is the Gauge-Higgs part of the action and $S_v$ is a potential term for the Higgs field [37], which introduces in leading order quadratic and quartic self couplings for the Higgs field. The overall coupling, $g_3^2$, is related to the temperature and the running coupling of the four dimensional theory.

The string tension in this effective theory will, of course, be non-vanishing. Its dependence on $g_3$ and the coupling constants in $S_v$, however, might be complicated. The analysis of the spatial string tension in (3+1)-dimensional QCD at finite temperature thus will yield crucial information about the structure of the effective theory deduced in the framework of dimensional reduction. In fact, the early simulations suggested that the spatial string tension would be temperature independent or only weakly dependent on temperature. This would suggest a rather complicated structure for the dimensionally reduced theory where the Higgs sector and non-static modes would substantially influence the spatial string tension. Only recently spatial Wilson loops have been studied in the pure $SU(2)$ gauge theory at temperatures substantially larger than the critical temperature. Here it turns out that $\sigma_s$ rises rapidly with $T$ [34,35]. Somewhat surprisingly the detailed analysis on large lattices and close to the continuum limit shows that the spatial string tension at large temperatures is even numerically close to what has been found for the string tension in three dimensional $SU(2)$ gauge theory [39]. One finds for the (3+1)-dimensional $SU(2)$ gauge theory and temperatures $T > 2T_c$ [40],

$$\sqrt{\sigma_s(T)} \ = \ (0.369 \pm 0.014) g^2(T)T \ , \ T \gtrsim 2T_c \ . (9)$$

where $g^2(T)$ is the 2-loop form of the temperature dependent running coupling constant. Here the additional free parameter in the $\beta$-function is determined as $\Lambda_T = 0.0076(13)T_c$.

The spatial string tension has been extracted from pseudo potentials, $V_T(R)$, constructed from space-like Wilson loops of size $R\times S$

$$V_T(R) \;=\; \lim_{S\to\infty} \ln \frac{W(R,S)}{W(R,S+1)} \quad . \tag{10}$$

These pseudo potentials have been calculated on lattices of size $N_\tau\times 32^3$ at a gauge coupling, $\beta = 2.74$, which is known to be the critical coupling for the deconfinement phase transition on a lattice of size $16\times 32^3$ [38]. The lattices with temporal extent $N_\tau$ at this value of the gauge coupling thus correspond to temperatures $T/T_c = 16/N_\tau$. From the results for the pseudo potentials shown in Fig.7 it is obvious that the slope at large distances, $\kappa(T)$, rises rapidly with $T$. This slope is related to $\sigma_s$ through

$$\sqrt{\sigma_s(T)}/T \;=\; \sqrt{\kappa(T)}N_\tau \quad . \tag{11}$$

In Fig.7b we show the ratio $T/\sqrt{\sigma_s}$. It is obvious from this that $\sigma_s$ rises slower than linear in $T$. Note that the scale used for $T/T_c$ in this figure is logarithmic. The linear rise seen in this figure thus suggests a logarithmic dependence of $T/\sqrt{\sigma_s(T)}$ on $T/T_c$. In fact, the temperature dependence is well described as being proportional to the inverse running coupling constant $g^{-2}(T)$. A fit with such an ansatz yields the result given in Eq.9.

This can be compared with the string tension of the three dimensional $SU(2)$ gauge theory [39],

$$\sqrt{\sigma_3} = (0.3340\pm 0.0025)g_3^2 \quad . \tag{12}$$

In the gauge part of the effective theory obtained from dimensional reduction the coupling constant is given by $g_3^2 = g^2(T)T$. This result thus suggests that the spatial string tension is dominated by the pure gauge part of the effective three-dimensional theory. In fact, it has explicitly been checked that $\sigma_s$ is rather insensitive to the structure of the Higgs sector of $S_{eff}$ [41].

In the regime $T_c\le T < 2T_c$ the behaviour of $\sigma_s$ differs from the above scaling behaviour. A flux tube model [36,42] suggests that in this regime the rise of $\sigma_s$ with $T$ is due to the squeezing of the vibrating surface by the shorter temporal direction of the four dimensional lattice. The spatial string tension is then expected to behave as

$$\sigma_s(T) = \frac{T}{T_c}\sigma \quad , \quad T\le T_c \lesssim 2T_c \quad . \tag{13}$$

This seems to be roughly fulfilled by the data shown in Fig. 7b.

The general structure of $\sigma_s$ discussed above has also been observed in much simpler gauge models like the $Z(2)$ gauge model in (2+1) dimensions [36,42]. In this case one can also distinguish two regimes in the high temperature phase, in which $\sigma_s$ shows a different functional dependence. In the region between $T_c$ and $2T_c$ the rise in $\sigma_s$ can be accounted for by the squeezing of the chromo-electric flux tube due to the shorter temporal direction and is still characterized by the parameters of the four dimensional theory, while for $T > 2T_c$ fluctuations are suppressed in the temporal direction and the behaviour of the string tension is controlled by the three dimensional effective theory.

Although the high temperature behaviour of the spatial string tension seems to be quite naturally related to the temperature dependence of the pure gauge part of the effective three dimensional theory given by Eq. 8, the result is not entirely satisfactory. The temperature dependent running coupling constant, $g^2(T)$, extracted from the fits to $\sigma_s$ turns out to be nearly a factor of two larger than the running coupling constant extracted from the heavy quark potential [43]. This alone is not disturbing, as different observables may well lead to different values for the $\Lambda_T$-parameter appearing in $g^2(T)$ as long as the $O(g^4)$ corrections are not fixed. The fact that both quantities seem to be described by the effective action, Eq. 8, with different choices of $g^2(T)$ is, however, unsatisfactory. It has been argued that the spatial string tension is not defined through a static operator and a comparison with the string tension of the effective theory thus requires the inclusion of additional contributions from non-static parts of the Wilson loop operator [37,43]. In this case, however, it would be entirely unclear why the proportionality factor found for the spatial string tension is that close to the va-lue found for the three dimensional theory. This point clearly requires further investigations.

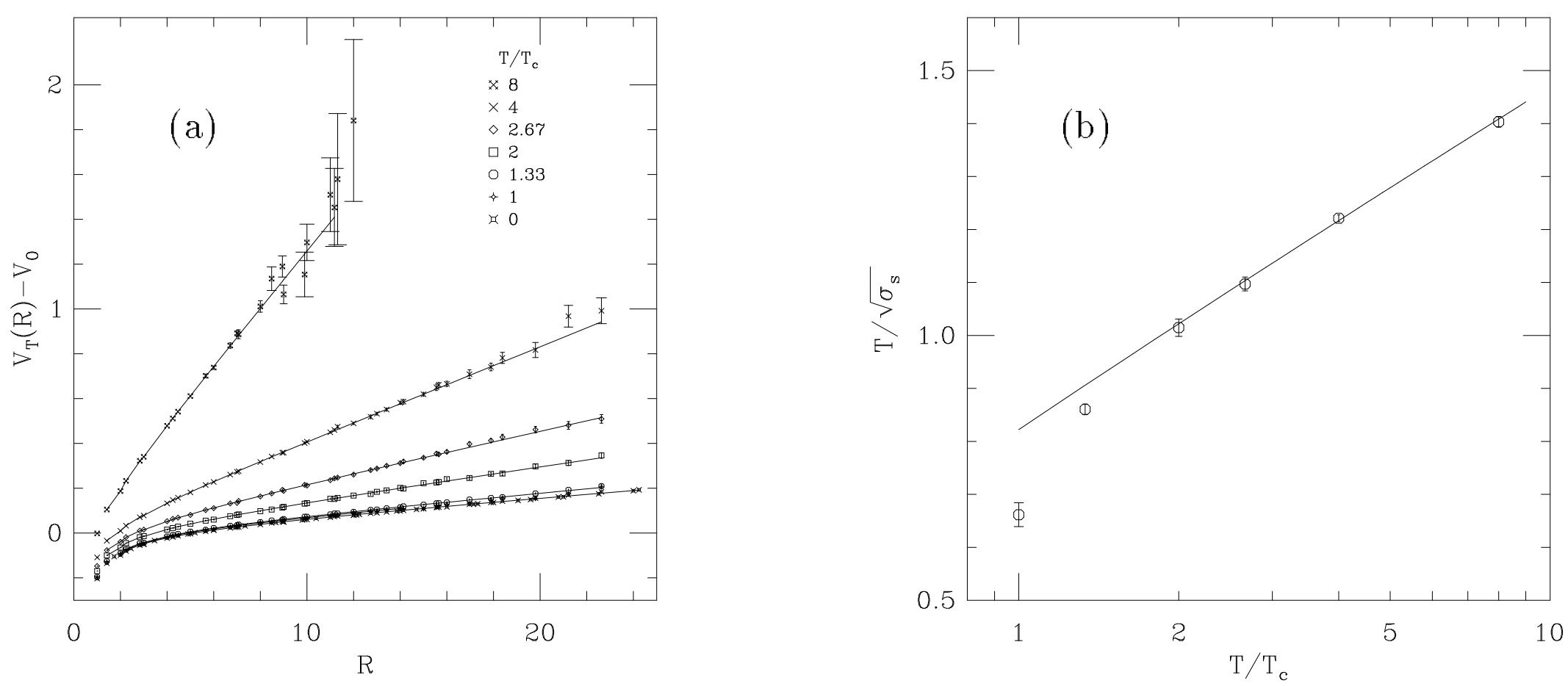

Figure 7. Temperature dependence of the pseudo-potentials $V_T(R)$ minus the constant self-energy contribution $V_0$ on lattices of size $N_\tau \times 32^3$ for $\beta = 2.74$ versus the spatial separation $R$ measured in lattice units (a). A fit to the linear part of these potentials yields the spatial string tension. The inverse square root of the spatial string tension in units of the temperature is shown in (b).

### 3.2. Hadron Correlators

During the past few years the temperature dependence of correlation functions with hadronic quantum numbers has been studied intensively. Here attention has focused on the behaviour of two-point functions at large spatial separations. The aim has been to extract from the exponential fall-off of these correlators screening masses, which should yield information about the existence or non-existence of bound states with the quantum numbers selected by the correlation function used. In the baryon and (pseudo-)vector meson channels these screening masses turned out to be compatible with the propagation of nearly free quarks, possibly with some residual interaction among them. One finds

$$m_{\rm Hadron} \simeq m_{\rm free} \simeq n\pi T \quad , \tag{14}$$

with $n = 2$ for mesons and $n = 3$ for baryons, i.e. the screening masses are close to multiples of the lowest Matsubara frequency, $p_{0,min} = \pi T$ for free fermions, which characterizes the exponential fall-off of a fermion correlation function at large spatial separations.

The (pseudo-)scalar sector, however, shows a markedly different behaviour. Here the screening masses are only about half as large as expected from the propagation of two independent quarks. This might indicate that bound states with the quantum numbers of the pion still exist in the QCD plasma phase. A two-particle bosonic bound state would have a vanishing lowest Matsubara frequency and thus could have a smaller screening mass. Some results for screening masses in four flavour QCD obtained from spatial hadron correlation functions is summarized in Fig.8. Similar results have been obtained for two flavour as well as quenched QCD.

In order to clarify the consequences for the excitation spectrum of QCD it is necessary to further study the analytic structure of the hadronic two-point functions. Recently the behaviour of these correlators at temporal separation [28,29] as well as their sensitivity to changes in the boundary conditions for the fermion fields has been studied [30]. Quite a different behaviour is expected in these cases as the screening masses will not be influenced by the non-vanishing Matsubara frequency. However, new difficulties in the interpretation of the correlation functions will show up.

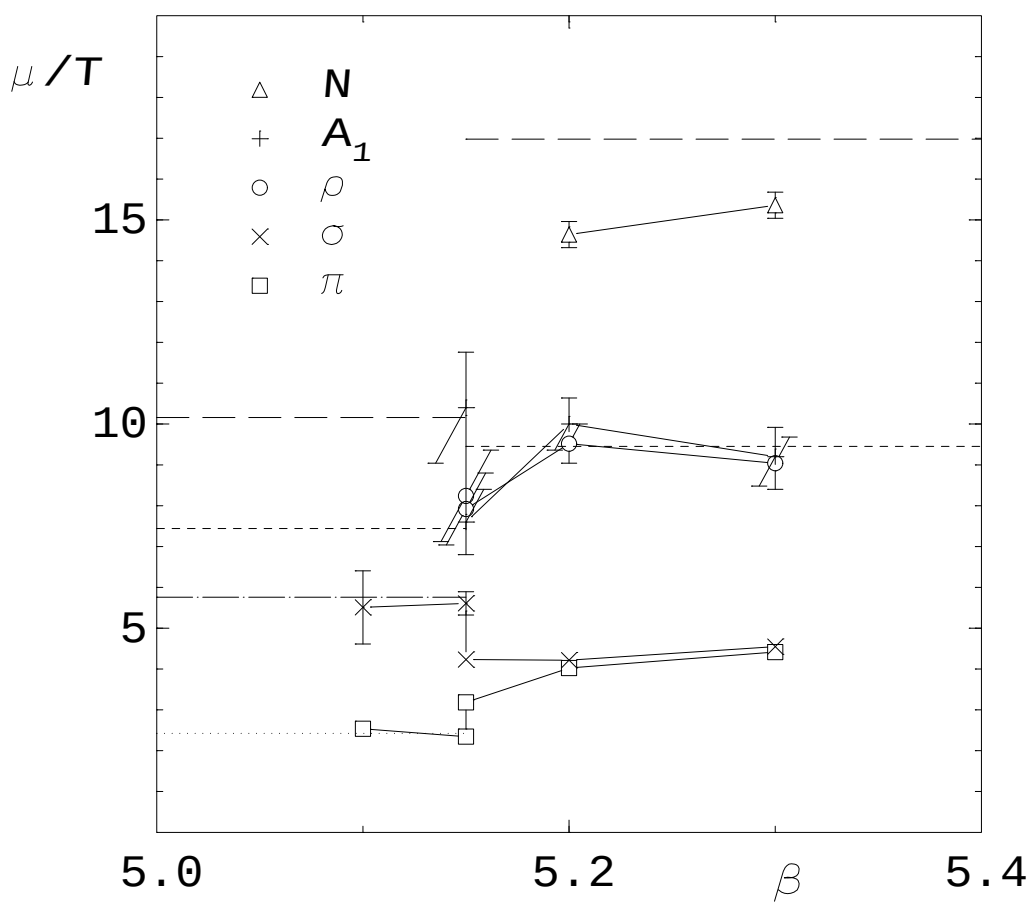

Figure 8. Hadronic screening length in four-flavour QCD determined from spatial correlation functions on lattices of size $8 \times 16^3$.

At non-zero temperature hadronic correlation functions can only be studied in a short Euclidean time intervall, $0 \le \tau \le 1/2T$. As a consequence, higher momentum excitations will contribute to correlation functions, in which source and sink are seperated in the temporal direction. This will strongly affect the shape of the correlation function at all accessible distances, which can be seen explicitly in the case of a meson correlation function, constructed out of a quark and an anti-quark operator. To be specific we will in the following discuss meson correlation functions on Euclidean lattices of size $N_\tau \times N_\sigma^3$ constructed within the staggered fermion formulation of lattice QCD. In particular we will describe numerical results obtained from simulations of four-flavour QCD using dynamical quarks of mass $ma = 0.01$ on lattices of size $8 \times 16^3$ [29]. The meson two-point functions can be constructed using local meson operators, $H(x)$, as sources. These are defined in terms of quark and anti-quark sources, $\chi(x)$ and $\bar{\chi}(x)$, as

$$H(x) \;=\; \phi_H(x)\bar{\chi}(x)\chi(x) \tag{15}$$

where $\phi_H$ is the phase factor, which depends on the quantum numbers of $H$. We use correlation functions with separations in the $x_0$-direction (temporal direction), which project onto a fixed momentum, $p$, of the meson,

$$G_H(x_0,p) = \frac{1}{N^3} \sum_{\substack{x= \\ (x_1,x_2,x_3)}} e^{ipx} \, \langle H(x_0,x)H(0,0)\rangle \, . \tag{16}$$

The correlation function, $G_H$, will receive contributions from quark fields carrying arbitrary momenta, subject to the constraint that their sum is equal to the fixed external meson momentum. In the case of non-interacting quarks and anti-quarks the correlation function can be calculated explicitly. In particular one finds for vanishing external meson momentum ($p = 0$) on an infinite lattice

$$G_H(x_0,0) = \frac{96T^3}{\pi^2} \int_0^\infty dy \frac{y^2 \cosh^2(y(2Tx_0-1))}{\cosh^2 y} \, . \tag{17}$$

This correlation function, together with results on finite lattices with staggered fermions is shown in Fig.9 on a logarithmic scale. The deviation from a simple exponential decay, which one would expect for a free meson correlation function, is clearly visible. In particular at short distances the high momentum modes from the free motion of the constituent quarks give rise to a steep rise of the correlation function.

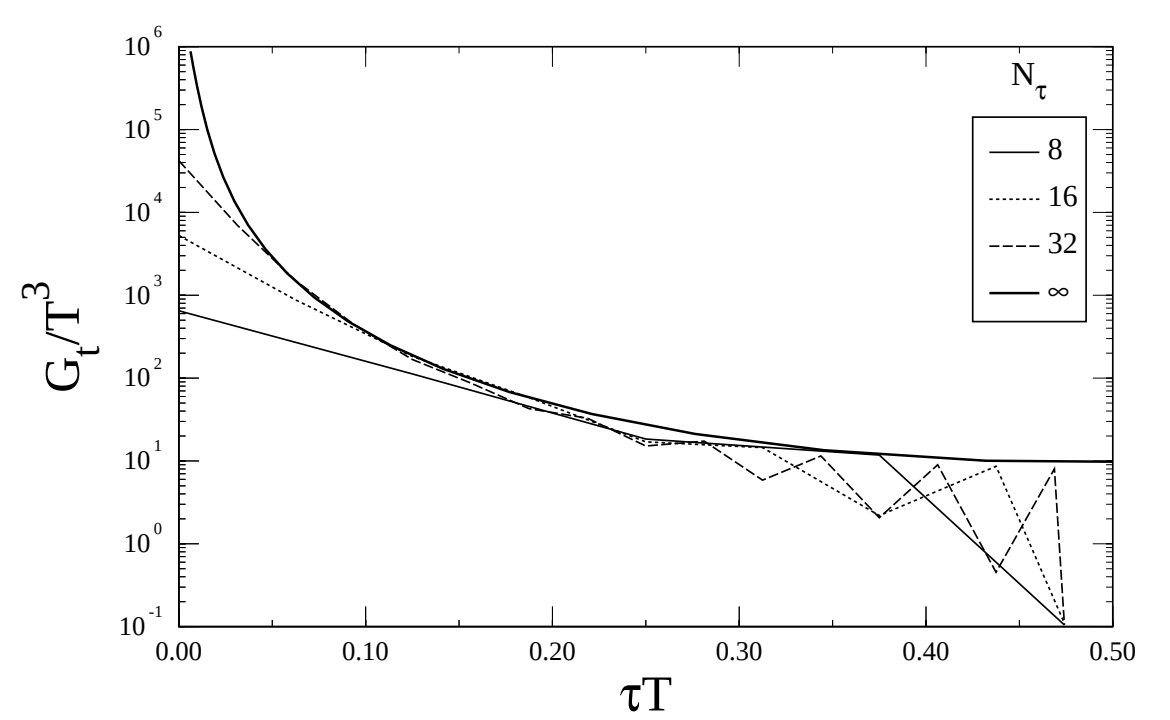

Figure 9. Meson correlation functions in the high temperature limit (lowest order perturbation theory) on lattices with infinite spatial extent and various values of $N_\tau$. Also shown is the result in the continuum limit ($N_\tau \to \infty$) given by Eq. 17.

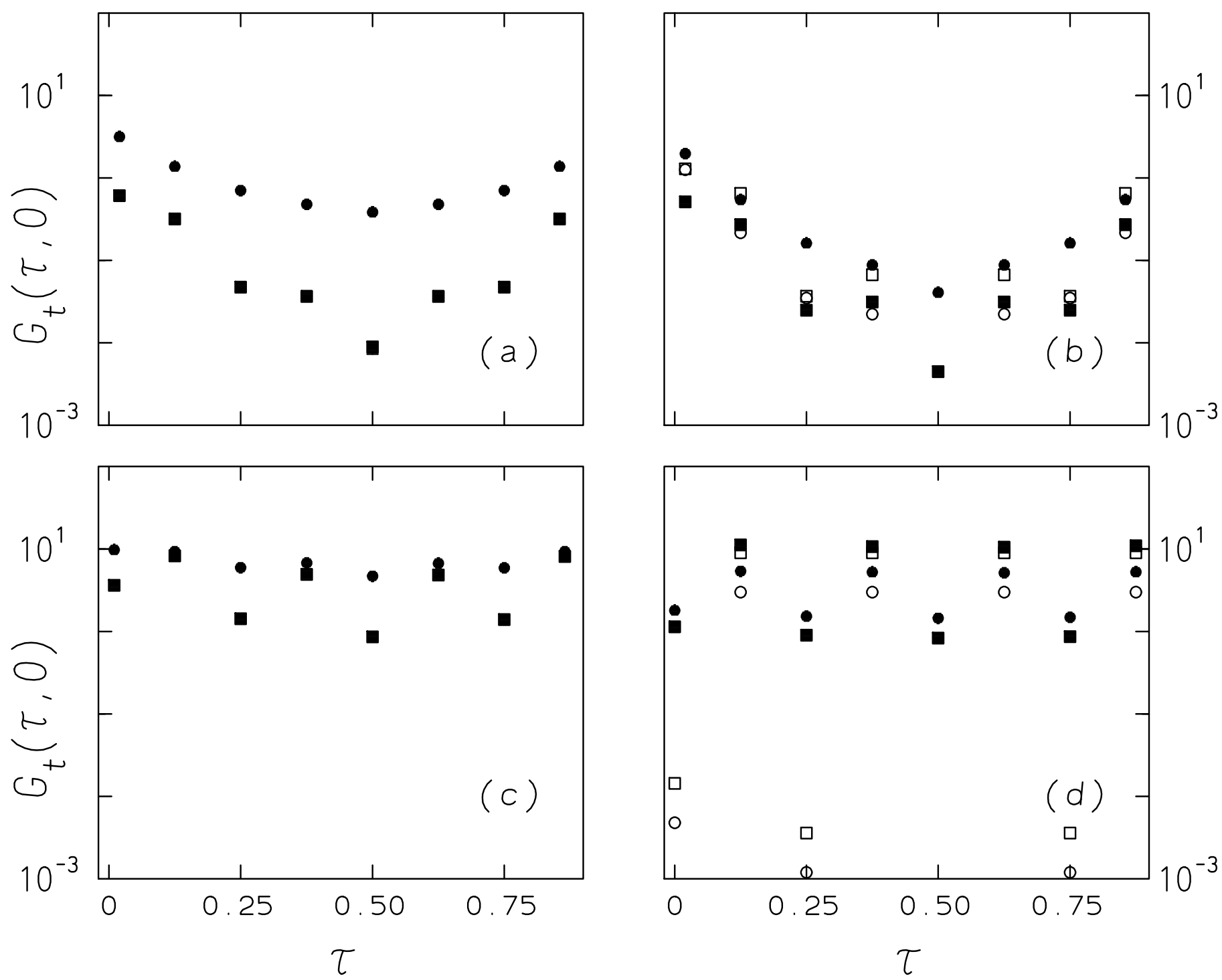

Figure 10. Pseudo-scalar (dots) and vector meson (squares) correlation functions constructed from point (upper row) and wall source operators (lower row) at two values of the gauge coupling in the QGP phase. Fig. (a) and (c) are for $6/g^2 = 5.3$ whereas Fig. (b) and (d) correspond to $6/g^2 = 6.5$. The open symbols denote the leading order perturbative result.

In order to become sensitive to low momentum excitations in temporal correlation functions, which would allow one to determine the lowest mass in this quantum number channel, one either has to perform fits with several exponential functions [28] or one has to construct correlation functions which project dominantly onto low momentum excitations. The latter approach has been followed in studies of the hadron spectrum at zero temperature, where one uses wall source operators,

$$H_w(x_0, x) = \frac{8}{N^3} \sum_{e,e'} \phi(x) \overline{\chi}(x_0, x+e) \chi(x_0, x+e'). \quad (18)$$

Here $e$ and $e'$ denote all even sites in the 3-dimensional hyper-plane of the lattice. The perturbative analysis of correlation functions with wall source operators shows that they only receive contributions from terms with small internal fermion momenta. In Fig.10 we show results for temporal correlation functions using point sources as well as wall sources. One clearly sees that with increasing value of the gauge coupling (increasing temperature) the wall source correlation functions become flatter while the point source correlation functions become steeper. The latter reflects the importance of high momentum excitations in the point-point correlation functions. These higher excitations completely dominate the fall-off of the correlation functions at moderate distances and give rise to large effective masses. The wall source correlators on the other hand become sensitive to the low momentum excitations. In fact, the operator used in the correlation functions is constructed such that in the case of non-interacting constituent quark fields it pro-jects onto the state where the constituent quarks carry vanishing momentum.

In Fig.11 we show results for the temporal masses extracted from the wall source correlation functions and compare them with twice the effective quark mass. As can be seen the temporal mass in the vector channel drops rapidly above $T_c$ and is compatible with twice the effective quark mass. In the pseudo-scalar channel the temporal mass also remains small. In fact, for $T \gtrsim T_c$ it still is smaller than twice the effective quark mass, which may indicate the existence of attractive interactions in this channel.

The results discussed here suggest that there are no bound states with large masses in the high temperature phase. In particular the screening masses extracted from the temporal correlation functions in the pseudo-scalar channel are a factor two smaller than those found from the spatial correlators. This suggests that also in the pseudo-scalar channel the non-vanishing Matsubara frequencies for fermions strongly influences the screening mass in the pseudo-scalar channel. This gets further support from the analysis of spatial correlators in which the quark fields are forced to obey periodic boundary conditions in the temporal direction. Above $T_c$ the screening masses extracted from such correlators are significantly smaller than those extracted from the corresponding operator with periodic boundary conditions for the fermion fields [29,30]. Below $T_c$, however, both type of correlation functions yield identical values for the screening masses. This indicates that the fundamental excitations are well localized bosons in the low temperature phase. In the high temperature phase the fermionic substructure becomes visible.

### 3.3. Pseudo Wave Functions

Another set of spatial correlation functions, whose structure has been examined during the past year are the so called *spatial wave functions* [44,45]. Similar to the spatial hadron correlation functions, these correlators are not directly related to the wave functions of hadrons. However, the fact that they still show an exponential decay [44], which is not the case in leading order perturbation theory, was again taken as indication for some non-perturbative effects in the high temperature phase. It has been suggested that the behaviour of these pseudo wave functions can also be understood in terms of the structure of the effective three dimensional theory for the high temperature phase [45]. The pseudo wave functions would then describe the bound states of heavy quarks with a mass given by Eq. 14 (with $n = 1$), which are bound by a confining potential with a temperature dependent string tension, which at high temperatures is given by Eq. 9. Such a scenario suggests that the pseudo wave functions drop exponentially like

$$|\psi(R)| \sim \exp\left(-\sqrt{\sigma_s(T)\pi T} R^{3/2}\right) \quad . \tag{19}$$

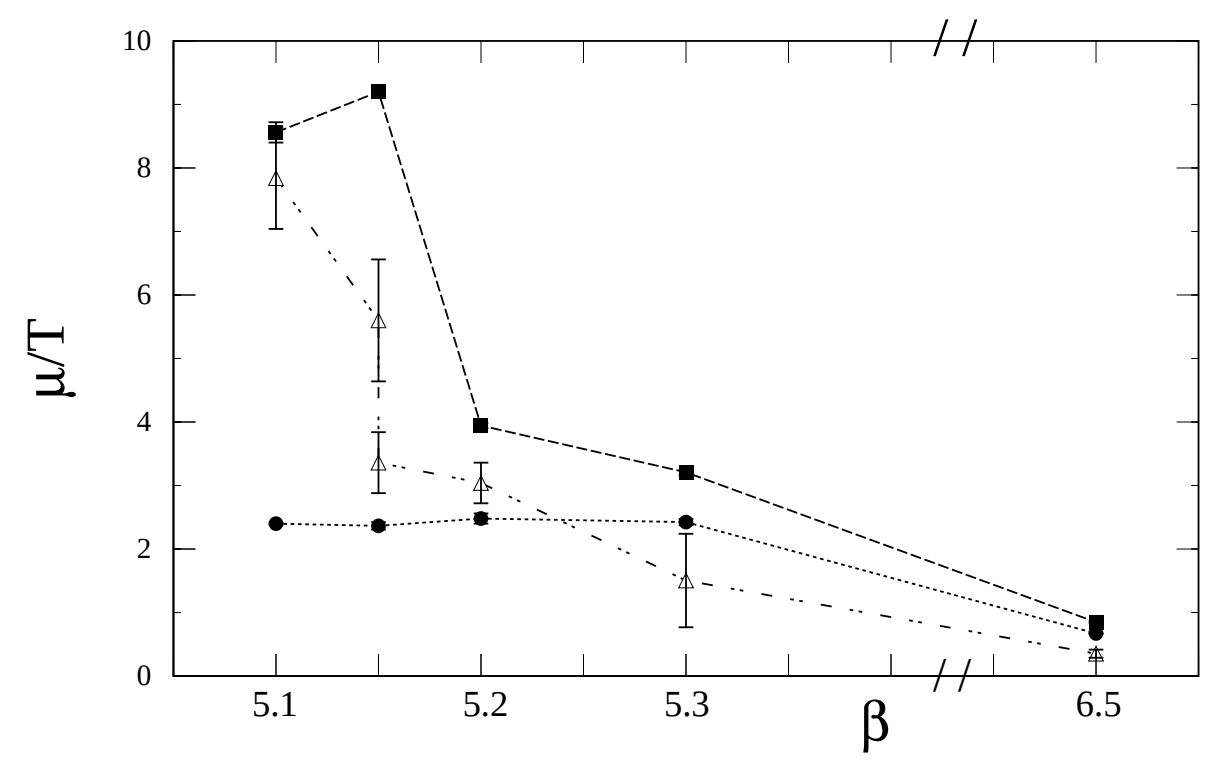

Figure 11. Temporal screening masses in units of the temperature versus the gauge coupling $\beta = 6/g^2$ for vector (squares) and pseudo-scalar (dots) mesons. Also shown is twice the effective quark mass in Landau gauge (triangles). The results shown are for four-flavour QCD on a lattice of size $8 \times 16^3$. The critical coupling for the chiral phase transition on this size lattice is $\beta_c = 5.15$.

The now existing quantitative results for the spatial string tension at high temperatures as well as the available information on the spatial screening masses will allow one to check the relation between the pseudo wave functions, the hadronic screening length and the spatial string tension in detail [46].

## 4. CONCLUSIONS

In this talk we did not pay much attention to studies of thermodynamic observables like the transition temperature, the Debye screening mass or the temperature dependence of the equation of state. Of course, these quantities are of great importance for the phenomenology of the QGP and of direct relevance for the discussion of possible signatures for QGP formation in heavy ion experiments. The current status of numerical studies of theses observables has been discussed in detail in various reviews [1,47]. Certainly in these cases as well, we cannot be satisfied with the accuracy of presently existing results from lattice calculations and more detailed studies of these quantitites will clearly be performed in the future.

In this talk we tried to concentrate on some of the important conceptual questions in finite temperature QCD, which have to be answered with the help of numerical simulations. We have discussed the status of our understanding of the phase transition in two flavour lattice QCD and the temperature dependence of various correlation functions in the high temperature phase of QCD. In the first case it is of great importance to reach a coherent description of the QCD phase transition in both commonly used fermion formulations. We thus have to pursue detailed investigations of the phase structure of both theories, although the calculations, which currently are performed close to the strong coupling regime, may not yet be of direct relevance for continuum physics. Similarly in the second case we have to study in more detail the analytic structure of correlation functions, which are not directly related to physical observables, with the aim to arrive at a consistent description of the excitation spectrum of the high temperature phase QCD.

We have, so far, not obtained a definite anwer from lattice simulations for any of these problems. Most likely we will have to wait for computers in the Teraflop range in order to come closer to a satisfactory answer. However, even the partial answers we have obtained so far are encouraging and important for the preparation of further investigations along this line.

## 5. ACKNOWLEDGEMENTS

I would like to thank all my collaborators and colleagues in Bielefeld, G. Boyd, J. Engels, E. Laermann, T. Neuhaus and B. Petersson for many helpful discussions. I am also very grateful to N. Christ, S. Gottlieb, U. Heller, K. Kanaya, R. Sugar and D. Toussaint for guidance through their data collections on the two flavour phase transition with staggered as well as Wilson fermions. This talk has been prepared at the ITP workshop on *Finite Temperature QCD* in Santa Barbara. I would like to thank the Institute for Theoretical Physics at UCSB and in particular J. Kapusta and E. Shuryak for the kind hospitality extended to me. This work was supported in part by the National Science Foundation under Grant No. PHY89-04035.